\documentclass{elsart}

\usepackage{natbib}

\usepackage{epsfig}

\usepackage{amssymb}

\begin{document}

\def\etal{{\it et~al.}}

\begin{frontmatter}



\title{The population of GRB hosts}


\author[1]{N. R. Tanvir}

\address[1]{Department of Physics and Astronomy, University of Leicester,
Leicester, LE1 7RH.  United Kingdom}

\author[2]{A. J. Levan}

\address[2]{Department of Physics, University of Warwick,
Coventry, CV4 7AL.  United Kingdom}

\begin{abstract}
The properties of their hosts provide important
clues to the progenitors of different classes of 
gamma-ray bursts (GRBs).  The hosts themselves also constitute a
sample of high-redshift star-forming 
galaxies which, unlike most other methods, 
is not selected on the luminosities
of the galaxies themselves.  
We discuss what we have learnt from and about GRB
host galaxies to date.
\end{abstract}


\end{frontmatter}

\section{Introduction}
\label{intro}

Pinpointing the first long-duration GRB afterglows quickly resolved
the debate over their distance scale \citep{jvp97,Metzger97}, and the realisation
that their hosts were high-redshift star-forming galaxies
was one of the first pieces of evidence suggesting their progenitors
were massive stars \citep{Bohdan98}.

Subsequently, the characteristics of their (likely) host
galaxies have formed an equally important line of argument
regarding the nature of the short-duration bursts \citep{Gehrels05,Hjorth05,Bloom06}.

In addition to helping understand GRBs themselves,
hosts are becoming increasingly important as
a high-redshift population selected only 
by its star forming properties, and not dependent
on the luminosity of individual galaxies, which
is the case for most other samples.

In this contribution we discuss the latest developments
in our understanding of the population of GRB hosts, and consider likely
future directions.


\section{Long-duration bursts}
\label{longs}


As mentioned above, the fact that no long-duration bursts (LGRBs)
have been found in early-type galaxies was a strong
argument in favour of their association with massive star
death.
The question naturally arises exactly what properties
a star must have in order to produce a GRB at
the end of its life.  This is related to the important 
question of whether the properties, rate and/or
luminosity, of GRBs depends on the characteristics
of the stellar populations which produce them.
In particular, if GRB properties depend on the
chemical makeup of their progenitor stars, or
other aspects of its galactic environment, that will
influence the statistical properties of the sample
of host galaxies they select.

\subsection{GRBs and metallicity}

In the popular ``collapsar" model for the production
of GRBs, it has been argued that high (around solar
and above) metallicity single
Wolf-Rayet stars will lose too much mass and angular
momentum to produce the rapidly rotating massive
cores that ultimately collapse to produce GRBs \citep{Heger03}.

A number of observational studies are consistent
with the idea that GRBs are preferentially produced
by stellar populations that are at least moderately
poor (sub-solar) in heavy metals.  \citet{Fynbo03} first noted
that the high proportion of GRB host galaxies
above redshift $z\approx2$ that show Lyman-$\alpha$ in emission
suggests they are low dust, low metal
systems.  Subsequently, studies of samples of
GRB host galaxies in the mid- and far-IR and
submm \citep{Berger03,Tanvir04,Lefloch06,CastroCeron06}
have found fewer hosts are bright in these
bands than expected, given the large amount of
obscured star formation expected to be taking
place in such galaxies.  Again, a preference
for lower metallicities would help explain
this finding if the very high star-formation
rate galaxies correspond to higher metallicity
systems as is thought.  Note there is potentially
a selection effect here, since some GRB hosts
will not be identified in the first place if the optical
light of the GRB is extincted by dust.  However, 
several such ``dark bursts" have sufficiently good
positions to identify their hosts, and were included
in the samples studied \citep[eg.][]{Barnard03}.
A similar argument has been made for the 
five lowest redshift GRBs, which are all in rather
small, metal-poor  galaxies when 
compared to the population of low-$z$
star-forming galaxies \citep{Stanek06}.

Most recently, \citet{Fruchter06} compared the
host galaxies of GRBs and core-collapse supernovae
in a similar redshift range, roughly $z=0.5$--1
(the average and spread of redshift was also 
similar between the two).  
The characteristics
of the galaxies and also the positions of the exploding
stars, differed significantly between the samples.
In figures 1 and 2 we show somewhat updated 
mosaics of the GRB and SNe hosts respectively.
The supernova hosts are clearly more likely to be brighter,
frequently grand-design spirals, while the GRB hosts are
typically smaller and have
irregular/merger morphologies \citep[see also][]{Conselice05,Wainwright07}.
This could also be explained if the GRBs are preferentially
formed from lower-metallicity core-collapse supernovae.
A bias against finding GRBs enshrouded in dusty systems
should be more than matched by the same bias against
finding supernovae hidden by dust (recalling that GRBs can
burn through significant columns of intervening dust and so
may sometimes be found optically even when 
enshrouded \citep[eg.][]{Waxman00}).
However, \citet{Wolf07} argue that the same data, whilst
compatible with a mild metallicity dependence of GRB
rate/luminosity, would not be consistent with a
strong effect.

\begin{figure*}
\centerline{
\psfig{figure=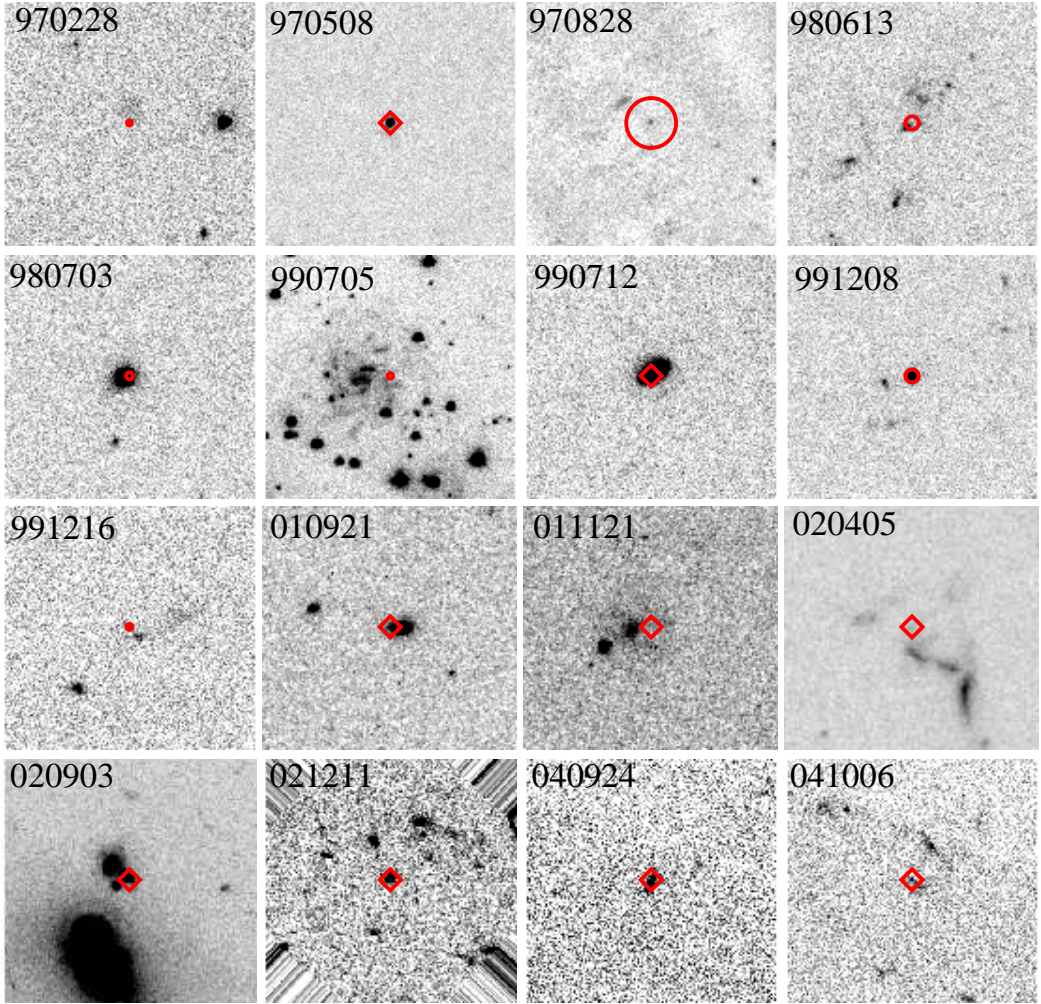,angle=0,width=14cm}}
\caption{A mosaic of HST images, each 7.5 arcsec on a side,
of the host galaxies of long-duration
GRBs with $0.3<z<1.0$.  Circles are 3$\sigma$ positional
uncertainties on the GRB positions.  Where the positions are
very well determined
(3$\sigma$ error less than 0.05 arcsec) the position is shown by a diamond.}
\end{figure*}

\begin{figure*}
\centerline{
\psfig{figure=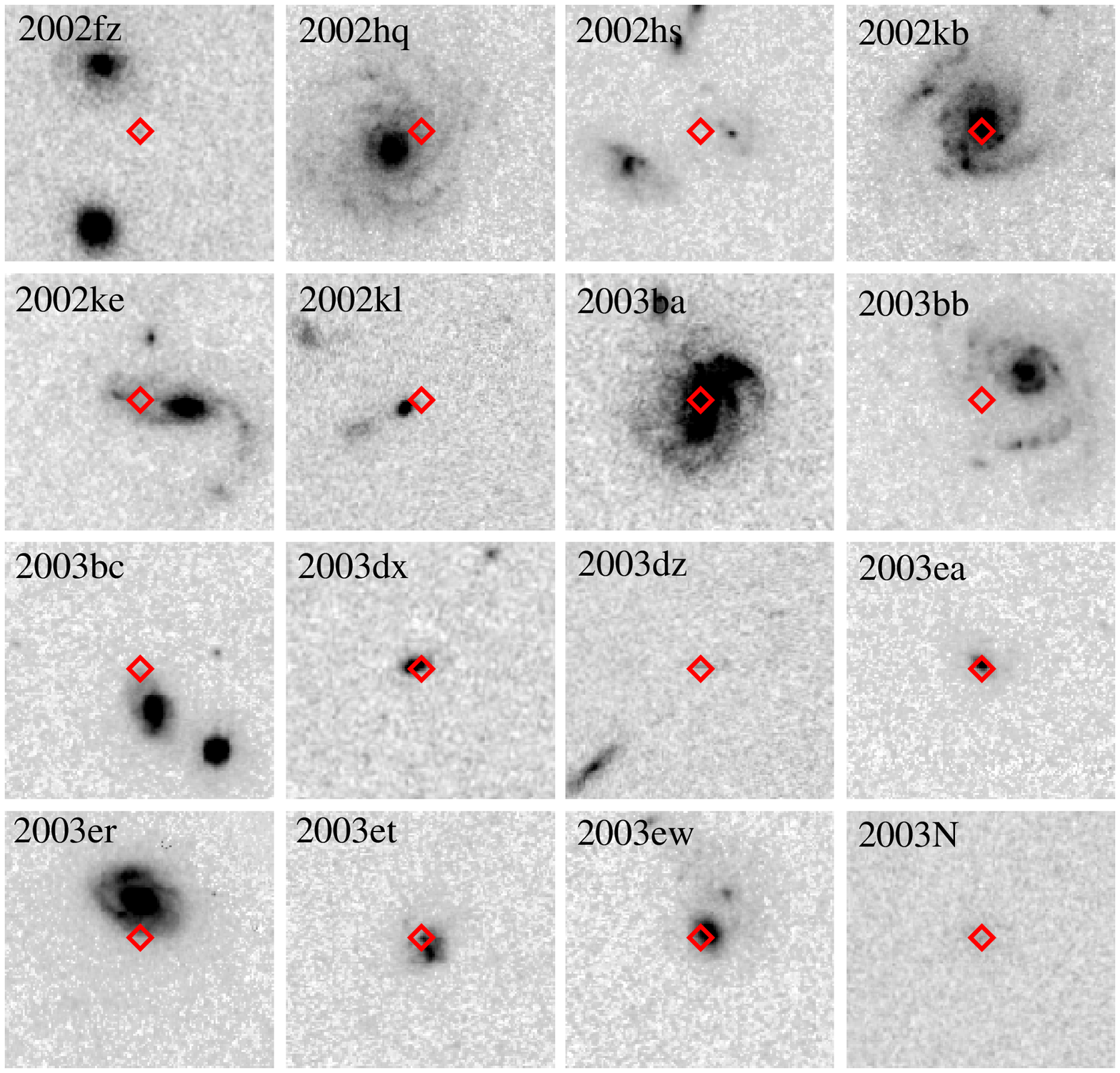,angle=0,width=14cm}}
\caption{As with figure 1, but for the hosts of core-collapse
supernovae found in the GOODS survey in the same redshift
range as the GRBs.  The difference
between the GRB and SN samples is clear to the eye, with
large and grand-design galaxies being more less common
amongst the GRB hosts \citep[see][]{Fruchter06}.
}
\end{figure*}

The chemical abundances of gas in the hosts of GRBs
can also be estimated directly via absorption line 
spectroscopy of the GRB afterglow itself.
Again, there could be a bias against higher 
metallicity, dusty galaxies, since the afterglow
must be optically bright to perform this analysis.
However, when such abundances have been determined
they show a wide range from about 1\% solar to
nearly solar \citep[eg.][]{Vreeswijk04,Prochaska07}.

\subsection{GRB host samples}

The immense luminosity of GRBs means that they can 
be detected in principle to very high redshifts.
Thus they can be used to select and characterise
galaxies from very early times up to the present.

If the rate of GRB production were the same for all
young stellar populations, then GRB host samples
should allow us to discriminate the proportions of
global star formation arising in different galaxy types,
and more generally map the history of star formation
in the universe \citep[eg.][]{Wijers98,Trentham02}.
As we have discussed above, it seems unlikely that GRBs
do trace star formation in a completely unbiased
way.
However, GRB selection will favour hosts with high star
formation rates (and probably, lower metallicities),
but otherwise is not biased against small faint
galaxies, which are typically missed in 
other flux-limited catalogues.  Redshifts, metallicities
and gas dynamics can be determined in many cases
from the afterglow spectroscopy.  A good example of
this power was GRB~020124, whose host was
undetected to $R\sim29.5$ in HST imaging \citep{Berger02}, but
was found to be a high column density DLA
at $z=3.2$ from the afterglow
\citep{Hjorth03}.

A number of attempts have been made to compare
GRB hosts as a whole to other high redshift populations.
For example, \citet{Jakobsson05} demonstrated that 
around $z\sim3$ the
bright end of the host luminosity function is
consistent with that expected by weighting by
star formation the Lyman-break galaxy luminosity function.

Many authors have noted that whilst occasional
bursts have been found in very red (ERO) star-forming
galaxies \citep[eg.][]{Levan06a,Berger07a}, the bulk of GRB hosts
are sub-L*, blue, low-dust, apparently young galaxies
with relatively strong line-emission and a high
specific rate of star formation \citep[eg.][]{Fruchter99,Lefloch03,Christensen04}.
Qualitatively these are similar characteristics to the
population of galaxies found in emission-line surveys for
Lyman-$\alpha$.  An interesting comparison is with the
wide area survey of \citet{Gawiser06}, for Ly-$\alpha$
emitters around $z\approx3.1$. In broad terms this population
is very like the GRB host sample in the same redshift range, albeit
that the Ly-$\alpha$ equivalent width is, unsurprisingly, somewhat
higher on the average.
Figure 3 shows the cumulative histograms of R-band continuum
luminosity (rest frame UV) for this sample together with
the published GRB hosts with $2.6<z<3.6$, illustrating their
similarity.

\begin{figure*}
\centerline{
\psfig{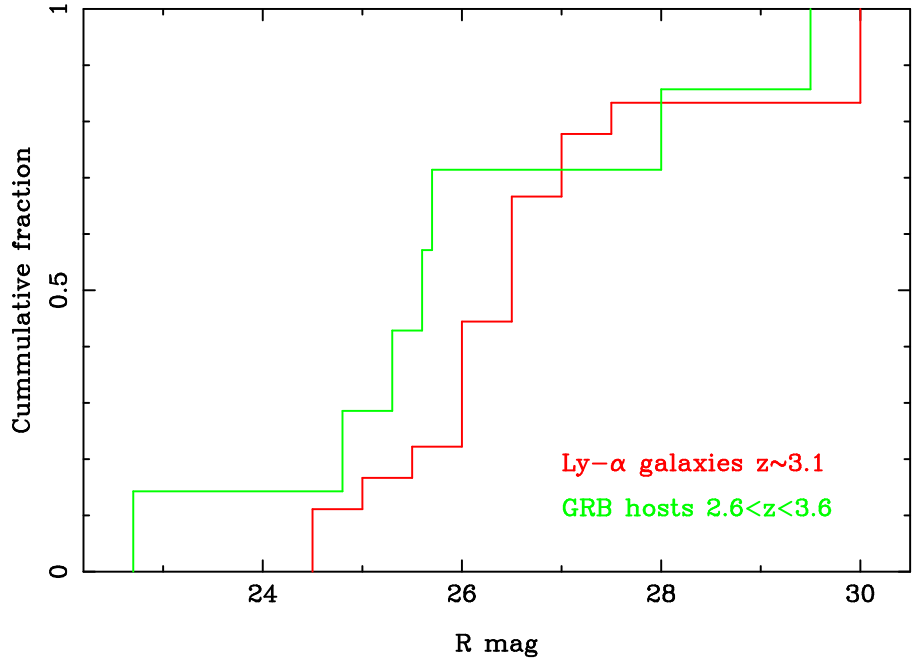}}
\caption{Cumulative luminosity histograms of 
broad band $R$ magnitudes for GRB hosts
and Lyman-$\alpha$ selected galaxies around $z\sim3$.
}
\end{figure*}

\section{Short-duration bursts}
\label{shorts}


The first few short-duration GRB afterglows seemed to 
paint a picture of being in galaxies at
redshifts of a few tenths and some of which
contained little or no young stellar population.
This was widely interpretted as being consistent 
with the neutron-star neutron-star (or neutron-star
black-hole) binary coalescence model for GRB
production.

Since then the picture has become murkier.  Several 
apparently short-duration GRBs have been found 
where the host is hard to identify, and most likely
is at much higher redshift $z>1$.  In particular,
GRB 060121 had a red afterglow and host
galaxy indicating a likely redshift $z>4$
and almost certainly $z>1.5$  \citep{Levan06b,AdUP06}.
The host and energetics of this burst are much
more typical of LGRBs, and the possibility remains
that it was actually a member of that class, despite the
short duration.

Although in many individual cases there can be
ambiguity over whether a given burst should be
in the short or long class \citep{Levan07b,Bloom07}, the weight of several 
likely high-z short bursts has led to speculation
that they form a separate sub-class \citep{Berger07b}.
It is worth commenting, though, that so-far all redshifts
for short bursts have come from their putative host rather
than the afterglow, and one consequence of a NS-NS
progenitor would be the possibility that the burst occurs
well beyond the optical extent of it's host, making 
definite association unclear in some cases.

\subsection{Short-duration bursts from nearby galaxies} 

In a parallel development \citet{Tanvir05} have shown that there
is a weak cross-correlation signal between the distribution
of BATSE short-duration bursts and galaxies in the
nearby universe.  In particular, they used the PSCz
galaxy redshift survey, which provides uniform selection
over 85\% of the sky, and found a positive signal
with the sample cut at various recession velocities out to
8000~km~s$^{-1}$ (approximately 110~Mpc).
Simulations suggested that this level of signal 
could be produced if between 10 and 25\%
of BATSE short bursts were coming from nearby
galaxies.

These on average must be considerably weaker bursts
than those found at cosmological distances.
The most likely progenitors are giant flares from
soft gamma-ray repeaters.  At least one such
flare from an SGR in the Milky Way (SGR 1806-20) was bright enough
that it could have been detected by BATSE to several tens of Mpc
\citep{Palmer05,Hurley05}.
In fact, only a very low rate of roughly one per
millenium per Milky-Way sized galaxy is sufficient
to explain the BATSE observed rate \citep{Levan07a,Ofek07}.

If of order 10\% of BATSE bursts were really from low
redshift galaxies it remains surprising that amongst those
well-localised by Swift and HETE-II there aren't any
clear-cut examples.  The best candidate is the weak burst
GRB 050906 whose BAT error circle contained the
outer parts of an actively star-forming galaxy
IC328 at a distance of about 130~Mpc \citep{Levan07a},
although the spectrum of the burst was significantly 
softer than previous giant flares.
Possibly the softer sensitivity of Swift/BAT and HETE-II compared
to BATSE makes it less likely that they will detect
SGR giant flares, which, on the basis of only three events,
seem to be typically hard (and thermal).

Interestingly, though, there are two candidates
for low-redshift short-duration bursts localised by
the Inter-Planetary Network (IPN).
Specifically, GRB 051103 was determined to have
occurred in a thin error region which lay close
to the outskirts of the galaxy M81 and at
that distance
the burst would have been quite consistent 
energetically with being an SGR
giant flare comparable to that from SGR 1806-20 \citep{Ofek06,Frederiks07}.
An even more compelling case may be GRB 070201, which
was found to overlap the outer part of the disk of
M31 \citep{Perley07,Hurley07}.  This was an extremely bright burst, and in that
regard, again, quite consistent with a very energetic SGR
flare at the distance of M31.  The only concern we might have is
that two such rare events should occur in neighbouring
large spiral galaxies (M31 and the Milky Way) 
within only two years of each 
other!

\section{Conclusions}
\label{conclusions}

The characteristics of their hosts has provided important clues
to the nature of GRB progenitors. 
Several lines of evidence suggest that LGRBs show some
preference for lower-metallicity hosts.
Particularly at high redshifts GRBs may be the
root to identifying and studying low-metallicity
star formation, and especially the faint end
of the galaxy luminosity function that is generally missed
in other surveys.
To fully realise the power of GRBs to select high-z
populations, it is important that statistical samples
of bursts and hosts with redshifts be as complete as
possible.  As it is, optical/nIR afterglows have been found
for nearly 80\% of well-positioned Swift LGRBs, but redshifts
for only about 50\% \citep{Tanvir07}.

Our understanding of the class of short-duration bursts
is at an earlier stage, but has seen huge progress in
the past two years.  Hosts have proved crucial to these
breakthroughs, not least because redshifts have
yet to be found directly for any short-burst
afterglow.  By way of illustration of the current,
rather confusing, state of affairs, we show in figure 4
a panel of hosts (or candidate hosts) of various short
duration GRBs, which range from the very nearby
candidates for SGR giant flare bursts, via the intermediate
redshift likely NS-NS progenitors, to the new ``population"
of apparently high redshift short bursts whose nature remains
controversial. 

\begin{figure*}
\centerline{
\psfig{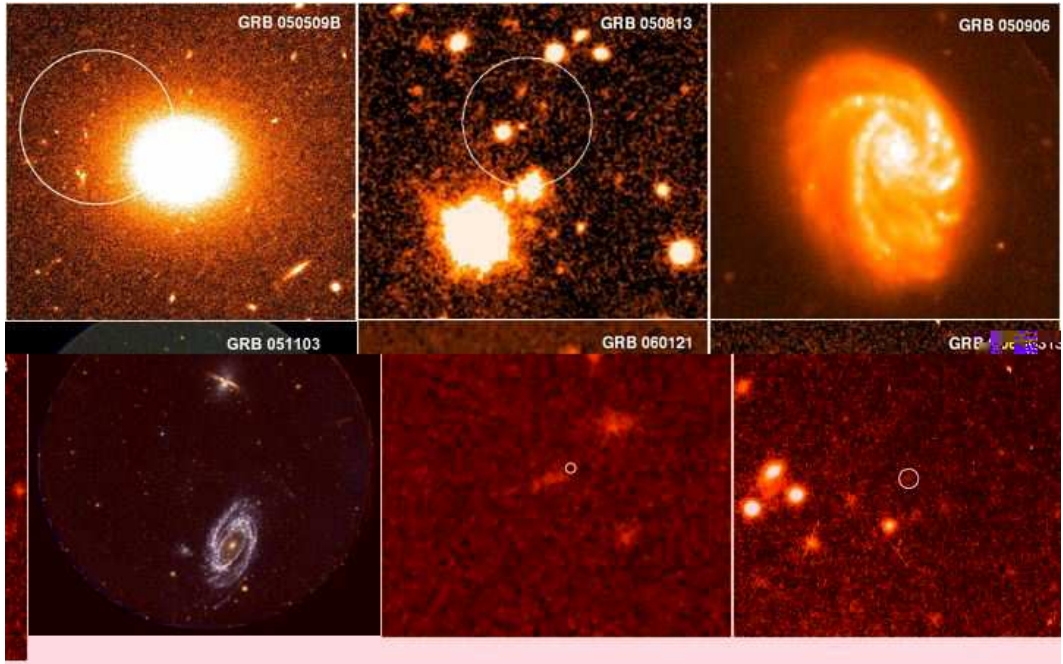}}
\caption{Panel showing various (candidate) host galaxies
of short-duration bursts.  This illustrates the surprising
diversity seen to-date, from the moderate-redshift,
high stellar mass galaxies expected to dominate 
for NS-NS progenitors, exemplified by GRB~050509B,
to possible low-redshift SGR giant
flare events (GRB~050906 and GRB~051103),
and the apparently very high redshift cases such
as GRB~050813, GRB~060121 and GRB~060313.
Note, the size of the images on the sky varies 
considerably in this panel.
}
\end{figure*}



\begin{thebibliography}{}


\bibitem[Barnard \etal(2003)]{Barnard03} Barnard V.~E. \etal, 2003, MNRAS, 338, 1 
\bibitem[Berger \etal(2002)]{Berger02} Berger E. \etal, 2002, ApJ, 581, 981 
\bibitem[Berger \etal (2003)]{Berger03} Berger E. \etal, 2003, ApJ, 588, 99
\bibitem[Berger \etal(2007a)]{Berger07a} 
Berger E., Fox D.~B., Kulkarni S.~R., Frail D.~A., Djorgovski S.~G., 2007, 
ApJ, 660, 504 
\bibitem[Berger \etal (2007b)]{Berger07b} Berger E. \etal, 2007, ApJ, 664, 1000
\bibitem[Bloom \etal (2006)]{Bloom06} Bloom J.~S. \etal, 2006, ApJ, 638, 354 
\bibitem[Bloom \etal (2007)]{Bloom07} Bloom J.~S. \etal, 2007, ApJ, 654, 878 
\bibitem[Castro Cer{\'o}n \etal(2006)]{CastroCeron06} Castro Cer{\'o}n J.~M., Micha{\l}owski 
M.~J., Hjorth J., Watson D., Fynbo J.~P.~U., Gorosabel J., 2006, ApJ, 653, 
L85 
\bibitem[Christensen, Hjorth, \& Gorosabel(2004)]{Christensen04} Christensen L., Hjorth J., Gorosabel 
J., 2004, A\&A, 425, 913 
\bibitem[Conselice \etal (2005)]{Conselice05} Conselice C.~J. \etal, 2005, ApJ, 633, 
29 
\bibitem[de Ugarte Postigo \etal (2006)]{AdUP06} de Ugarte Postigo A. \etal, 2006, ApJ, 
648, L83 
\bibitem[Frederiks \etal(2007)]{Frederiks07} Frederiks D.~D., Palshin V.~D., Aptekar 
R.~L., Golenetskii S.~V., Cline T.~L., Mazets E.~P., 2007, AstL, 33, 19 
\bibitem[Fruchter \etal(1999)]{Fruchter99} Fruchter A.~S. \etal, 1999, ApJ, 519, 
L13 
\bibitem[Fruchter \etal (2006)]{Fruchter06} Fruchter A.~S. \etal, 2006 Nature 441, 463
\bibitem[Fynbo \etal (2003)]{Fynbo03} Fynbo J.~P.~U. \etal, 2003, A\&A, 406, L63 
\bibitem[Galama \etal (1998)]{Galama98} Galama T.~J. \etal, 1998 Nature 395, 670 
\bibitem[Gawiser \etal(2006)]{Gawiser06} Gawiser E. \etal, 2006, ApJ, 642, L13 
\bibitem[Gehrels \etal (2005)]{Gehrels05} Gehrels N. \etal, 2005 Nature 437, 851
\bibitem[Heger \etal (2003)]{Heger03} Heger A., Fryer C.~L., Woosley S.~E., Langer N., Hartmann D.~H., 2003, ApJ, 591, 288 
\bibitem[Hjorth \etal(2003)]{Hjorth03} Hjorth J. \etal, 2003, ApJ, 597, 699 
\bibitem[Hjorth \etal(2005)]{Hjorth05} Hjorth J. \etal, 2005, ApJ, 630, L117 
\bibitem[Hurley \etal(2005)]{Hurley05} Hurley K. \etal, 2005, Nature, 434, 1098 
\bibitem[Hurley \etal (2007)]{Hurley07} Hurley K. \etal, 2007, GCN, 6103, 1 
\bibitem[Jakobsson \etal(2005)]{Jakobsson05} Jakobsson P., \etal, 2005, MNRAS, 362, 
245 
\bibitem[Le Floc'h \etal(2003)]{Lefloch03} Le Floc'h E. \etal, 2003, A\&A, 400, 499 
\bibitem[Le Floc'h \etal (2006)]{Lefloch06} Le Floc'h E., Charmandaris V., Forrest 
W.~J., Mirabel I.~F., Armus L., Devost D., 2006, ApJ, 642, 636 
\bibitem[Levan \etal(2006a)]{Levan06a} Levan A.~J. \etal, 2006a, ApJ, 647, 471 
\bibitem[Levan \etal (2006b)]{Levan06b} Levan, A.~J. \etal, 2006b, ApJ, 648, L9
\bibitem[Levan \etal (2007a)]{Levan07a} Levan A.~J. \etal, 2007a, MNRAS in press, arXiv:0705.1705 
\bibitem[Levan \etal (2007b)]{Levan07b} Levan A.~J. \etal, 2007b, MNRAS, 378, 1439 
\bibitem[Metzger \etal (1997)]{Metzger97} Metzger M.~R. \etal, Nature 387, 878
\bibitem[Ofek \etal(2006)]{Ofek06} Ofek E.~O., \etal, 2006, ApJ, 652, 507 
\bibitem[Ofek(2007)]{Ofek07} Ofek E.~O., 2007, ApJ, 659, 339 
\bibitem[Paczynski(1998)]{Bohdan98} Paczynski, B., 1998, ApJ, 494, L45
\bibitem[Palmer \etal(2005)]{Palmer05} Palmer D.~M. \etal, 2005, Nature, 434, 1107 
\bibitem[Perley \& Bloom(2007)]{Perley07} Perley D.~A., Bloom J.~S., 2007, GCN, 
6091, 1 
\bibitem[Prochaska \etal (2007)]{Prochaska07} Prochaska J.~X., Chen H.-W., 
Dessauges-Zavadsky M., Bloom J.~S., 2007, ApJ, 666, 267 
\bibitem[Stanek \etal (2006)]{Stanek06} Stanek K.~Z. \etal, 2006, AcA, 56, 333 
\bibitem[Tanvir \etal (2004)]{Tanvir04} Tanvir N.~R. \etal, 2004, MNRAS, 352, 1073
\bibitem[Tanvir \etal (2005)]{Tanvir05} Tanvir N.~R., Chapman R., Levan A.~J., Priddey R.~S., 2005, Nature, 438, 991
\bibitem[Tanvir \& Jakobsson(2007)]{Tanvir07} Tanvir N.~R., Jakobsson P., 2007, 
Phil. Trans of the Royal Society A, 365, 1377, arXiv:astro-ph/0701777 
\bibitem[Trentham, Ramirez-Ruiz, \& 
Blain(2002)]{Trentham02} Trentham N., Ramirez-Ruiz E., Blain 
A.~W., 2002, MNRAS, 334, 983 

\bibitem[van Paradijs \etal (1997)]{jvp97} van Paradijs J.~V. \etal, 1997, Nature, 386, 686
\bibitem[Vreeswijk \etal (2004)]{Vreeswijk04} Vreeswijk P.~M. \etal, 2004, A\&A, 419, 
927 
\bibitem[Wainwright, Berger, \& Penprase(2007)]{Wainwright07} Wainwright C., Berger E., Penprase 
B.~E., 2007, ApJ, 657, 367 
\bibitem[Waxman \& Draine(2000)]{Waxman00} Waxman E., Draine B.~T., 2000, ApJ, 
537, 796 
\bibitem[Wijers \etal(1998)]{Wijers98} 
Wijers R.~A.~M.~J., Bloom J.~S., Bagla J.~S., Natarajan P., 1998, MNRAS, 
294, L13 
\bibitem[Wolf \& Podsiadlowski(2007)]{Wolf07} Wolf C., Podsiadlowski P., 2007, 
MNRAS, 375, 1049 


%

\end{thebibliography}
\end{document}